# Mathematical Analysis of Dynamic Risk Default in Microfinance

1  Mohammed Kaicer

   Faculty of Sciences

   University of Ibn Tofail, Morocco Kingdom

2  Abdelilah Kaddar

   Faculty of Law, Economics and Social Sciences of Salé

   University of Mohammed 5, Morocco Kingdom

**Corresponding address:** Informatics System and Optimization Laboratory, University of Ibn Tofail, P.B 242, Kenitra, Morocco Kingdom

**e-mail of corresponding author:** mohammed.kaicer@uit.ac.ma

**Abstract:** In this work we will develop a new approach to solve the non repayment problem in microfinance due to the problem of asymmetric information. This approach is based on modeling and simulation of   ordinary differential systems where time remains a primordial component, they thus enable microfinance institutions to manage their risk portfolios by a prediction of numbers of solvent and insolvent borrowers ever a period, in order to define or redefine its development strategy, investment and management in an area, where the population is often poor and in need a mechanism of financial inclusion.

## 1.  Introduction

Microcredit is a major component of microfinance for the poor ($ 1.95 / person / day according to The World Bank 2017) a large part of this population does not have access to formal financial services, which almost excluded from the financial system, this is one of the main factors in the poor's inability to participate in the process of economic development of   regions where they live. The creation of microfinance industry, for over 30 years, aims to fight against poverty in improving the living conditions of the poor and allows them to integrate in the development of their region.

The amounts awarded as part of microcredit, which is a component of the microfinance, they take the form of financing and refinancing of income-generating micro projects where borrowers, with entrepreneurial capabilities, choose a type of credit agreement, individual loan contract and joint liability loan. The first is characterized by an interest rate higher than the group lending. In a group lending each member is required to repay the total amount of its members if they fail otherwise they will all have access to an individual lending if all members of the group succeed each loan cycle. This mechanism of co-responsibility replaces the absence of material guarantees; it aims to create social interaction through monitoring and mutual pressure and mutual help members. The problem of failure in the case of co-responsibility is that if at least one member fails this causes the failure of the entire group which prevents all members' access to future loans among all the Microfinance Institutions MFIs.

In the literature we find that the rate of reimbursement for group lending and the biggest individual loan carrying the risk of non-repayment in the collective case is due to the types of group members. A 'bad' borrower can negatively influence a 'good' borrower and vice versa. (Kaicer, 2014)

MFIs management requires a mastery of its portfolio borrowers in terms of performance and efficiency. The median return on equity for reporting institutions at the Mixmarket was 8.1% in 2016, against 7.9% in 2015 and 9.6% in 2014. The Non-Governmental Organizations NGOs in rural areas and banks argue that a positive change compared to the gross loan portfolio return aggregated to 26.5% in 2016, quite similar to 2015. Africa has had a return of 34.3% that is the highest in the world, while South Asia had the lowest return of 22.8% during the year (Table 1).

Global level the global risk portfolio (GRP) exceeding 30 days was 4.7% in 2016, slightly higher to 3.9% in 2015. The South Asia GRP, which was the lowest in 2015, increased to 1.9% in 2016, the largest change reported by MFIs. Africa MFIs have witnessed a deterioration in their portfolio as they recorded a PER of 6.9% in 2016.

The portfolio at risk is associated with the risk of non-repayment, closely related to the numbers of solvent and insolvent borrowers, this risk presents Achill's heel in microcredit and occupies a prominent place for better management of MFIs (Armendariz & Morduch, 2010).

This topic is also important for several academic researchers who proposed some approaches to diagnose and troubleshoot the risk of default. In terms of Modeling, risk is tackled by the application of approaches related to the exact sciences and the social sciences namely, game theory (Besley & Coate, 1995) (Tassel, 1999) applied to the theory of contracts (Tassel, 1999) (Ghatak, 1999)such mechanisms to refine loan contracts in the presence of asymmetric information regardless of the type of contract and type of borrowers.

The theory of incentive contracts (Maskin, 2011), (Bhole & Ogden, 2010) to minimize the effect of asymmetric information, like as optimization to minimize its effect of the asymmetry of information, optimized to maximize the welfare of the borrower under the budgetary constraints imposed by the MFI where it is proposed that the incentive take the form of refinancing the project in the event that the borrower fully or partially complies with the terms of his contract (Tedeschi, 2006). In addition to the aspect by data analysis (Kaicer, 2014) (Bourjade & Schindele, 2012)draw correlation structure by determining the characteristics of the solvent and borrowers who are insolvent. Adding predictive models to establish a credit scoring model. (Aboulaich, Khamlichi, & Kaicer, 2013). Other works were interested in best practices of microfinance (Armendariz & Morduch, 2010) the authors have prescribed successful international microfinance actors where the utility is manifested in the welfare of borrowers and the portfolio quality of MFIs. To design managerial mechanisms to better manage the risk of non-repayment which include the work of (Norell, 2001), (Milgram, 2001), (Park & Ren, January 2001), (Hulme & Mosley, 1996). However, the number of academic work remains low relative to research in quantitative finance (Kaicer, 2014).

The main idea of this paper is to propose a new approach to understand the manipulation of the risk of non-repayment and defend theoretically and practically for the minimization of the number of unpaid, indeed if the rate of unpaid increases over time precisely in a daily way, weekly, monthly or yearly, this leads to a risky portfolio and so it threatens the sustainability of MFIs (Kaicer & Aboulaich, 2014) The model illustrating this problem is a system of two non-linear differential equations.

Since enough time Systems of differential equations play an important role for modeling, the description and monitoring of dynamic phenomena, It should be noted that non-linearity is a necessary condition, but not sufficient to generate chaos or randomness, which are both due to asymmetric information. The modern concept of deterministic chaos is increasingly used in scientific contexts varying mathematics and physical dynamic systems and even complex temporal variations of all types economics (Kaddar, Sahbani, & Allaoui, 2018) , in chemistry, biology, physiology, sociology and even psychology. (Leiber, 1997)

Like any organization, the MFI tries to ensure its sustainability by minimizing the number of unpaid also by expanding its activities in other areas in order to ensure a broad expansion of its services. The approach envisaged in our contribution is to understand the dynamic phenomenon of non-repayment in a borrowing population according to the solvency criterion, this would allow MFIs to predict in the medium term as well as in the long term the numbers of solvents borrowers and those who are insolvent, even as the existing mutual impact between good borrowers and bad, regardless of contracts type offered by the MFIs, to implement new agencies and to withdraw from others where risk portfolios occur.

The approach in this paper is a process of a dynamic system corresponding to the adapted financial system with the microfinance environment and the services offered by its institutions. In our study the process is powered by input are probability distributions and who contribute in the evolution of the system see all figures attached Again, there are disruptions that are unobservable magnitudes by the MFI and affect the process (figure 1).

We will highlight the dynamics of the system, such as the evolution of borrowers according to their types Solvable / Insolvable, a system of differential equations noted $(SI)_t^{general}$ with the initial conditions $(S(0), I(0))$ of the general form:

$$(SI)_t^{geberal} \begin{cases} \dfrac{dS(t)}{dt} = f(S, I) \\ \dfrac{dI(t)}{dt} = g(S, I) \end{cases} \text{ with the initial conditions } (S(0), I(0)), \quad (1)$$

Where *f* and *g* are two differentiable functions in $(S(t), I(t))$ on $IR^2$.

## 2. The model

The proposed model addresses the dynamics of a borrower population where repayment of credits is a fundamental criterion for MFIs. In practice the mutual impact is strong between borrowers namely the socio-cultural environment and coexistence thus the influences are positive and negative in terms of compliance with the credit agreement.

Either a zone ℤ and let us consider a population of size *P*. Let *α* the proportion of a subpopulation ℤ' and who accepts the contracts offered by the MFI in ℤ so $(1-\alpha)P$ is still out of the microcredit system. In ℤ' we assume the existence of the solvent borrowers of size $\sigma\alpha P$ and insolvable of $(1-\sigma)\alpha P$. Let $\beta_1$ the proportion of insolvent that negatively influence solvent borrowers then they will not respect their contract of refinancing. Let $\beta_2$ the positive impact from the solvent to insolvent. *P* is practically composed by solvable that noted by *S*. and the insolvent by I. On the other hand, experience has shown that there is an exodus between the groups of solvent borrowers and insolvent in terms of credit cycles. This is modeled by two magnitudes (Kermack & McKendrick, 1927.):

$\mu_1 S$ (Respectively $\mu_2 I$) the number of solvent (or insolvent) borrowers who have left the credit cycle. For a period credit cycle *[0,t]*, the number of borrowers is a function of time noted by $S_t$ and $I_t$:

$$P_t = \frac{1}{\alpha}(S_t + I_t),\quad (2)$$

In MFIs, the performance of the portfolios is to know the variation inter-temporal between $S_t$ and $I_t$ relative to the same period. This diagnosis can be modeled by a system of differential equations. The reformulation of assumptions leads us to the evolution model of the number of solvent and insolvent borrowers over time, by a system of nonlinear differential equations noted $(SI)_t$. The following system determines under what insolvency conditions is manifested or not in a population in period $t$:

$$(SI)_t \begin{cases} \dfrac{dS_t}{dt} = \sigma \dfrac{1-\alpha}{\alpha}(S_t + I_t) - \beta_1 S_t I_t + \beta_2 S_t I_t - \mu_1 S_t \\ \dfrac{dI_t}{dt} = (1-\sigma)\dfrac{1-\alpha}{\alpha}(S_t + I_t) + \beta_1 S_t I_t - \beta_2 S_t I_t - \mu_2 I_t \end{cases} \quad (3)$$

Where $\beta_1$ represents the transmission coefficient of the compartment of the solvable to the compartment of insolvable by the negative effect of contact with insolvent and $\beta_2$ coefficient of transmission of the insolvency compartment to the credit by the positive effect of contact with solvable. Generally the weights can be seen as probability distributions, note that the model $(SI)_t$ presents a simplification of the complex system of the microfinance environment.

## 3. Numerical simulation and interpretations

In this section, we propose numerical simulations of the system $(SI)_t$. The results obtained give the evolution of $S_t$ and $I_t$ over time. Practically the parameters of the model are chosen in such a way as to summarize the missing information, and characterizing the socio-economic, geographic and demographic environment of the target area fixed by the MFI. These parameters can be estimated empirically based on field surveys on or on the data base borrowers integrated in IT of the organization wishing to study the problem of non repayment. As for our numerical resolution we adopt values between 0 and 1 since they are proportions or parameters of the laws of probability.

The test period is between 0 and 10 (x unit of time). For $\alpha = 0.1, \sigma = 0.5, \beta_1 = 0.1, \beta_2 = 0.4, \mu_1 = \mu_2 = 0$ is with initial population $(S, I) = (10000, 2000)$. A visualization of Figure 1 shows that, in the first of period, $S_t$ decreases to 8000 solvent borrowers then it becomes stable over the rest of the period. Quant à $I_t$ also decreases and reaches almost zero values throughout the period. Generally the graph shows that there is a concordance between $S_t$ and $I_t$. In this situation the MFI should diagnose the integration of individuals into microfinance systems in this area. Here, we are talking about financial inclusion and even the mutual influence between safe borrowers and risk borrowers $\beta_1 = 0.1, \beta_2 = 0.4$.

In figure 2, for $\alpha = 0.34, \sigma = 0.6, \beta_1 = 0.7, \beta_2 = 0.2, \mu_1 = 0.1$ with a strong migration to the microcredit system $\mu_2 = 0.9$,, the simulation of $(SI_t)$ generates an ideal situation for the MFI ($\beta_1 > \beta_2$). Indeed $S_t$ evolves exponentially throughout the period relative to a remarkable degradation of $I_t$ which reaches zero values from the beginning of the period.

So the strategy of the MFI, for development and management, should maintain as much as possible the values taken by $\alpha, \sigma, \beta_1, \beta_2, \mu_1, \mu_2$ and which characterize this area where the MFI offered (or will offer) its services or financing contract. The MFI is thus must strengthen its existing in this area and expect even more borrowers being out of the microcredit system and encourage them to incorporate (figure 3).

Othe important case for MFIs: on the period [0.10] with initial population from $(S, I) = (10000, 1865)$, $\alpha = 0.2, \sigma = 0.29, \beta_1 = 0.67, \beta_2 = 0.56, \mu_1 = 0.8$, *$\mu_2=0.41$*, and $S_t$ decreases towards a very small number of solvable, as for the number of insolvable, $I_t$ will reach values close to 0 from the beginning of the period and will grow from half the period. In this situation the portfolios at risk become remarkable and thus the sustainability of the MFI is strongly threat. Given this undesirable situation MFI should review its development strategy and management where he could withdraw from this area possibly characterized by the parameters $\alpha = 0.2, \sigma = 0.29, \beta_1 = 0.67, \beta_2 = 0.56, \mu_1 = 0.8$ et $\mu_2 = 0.41$. (figure 3)

In figure 4, we have as initial population $(S, I) = (10000, 20000)$ and $\alpha = 0.73, \sigma = 0.21, 0 \leq \beta_1, \beta_2 \leq 1, \mu_1 = 0.4$ and $\mu_2 = 0.9$. This situation remains the worst case for all MFIs, since at the beginning of the period one visualizes a fall acute of $S_t$ and $I_t$. $S_t$ converges to 0 and remains virtually zero throughout the period. Furthermore $I_t$ tends to 0 decreasing. Faced with this situation the MFI should absolutely react on the model parameter values to address this extreme fall.

4. **Conclusion and Discussion:**

The work done so far in the field of microfinance is trying to model the risk of default by the different disciplines of social sciences. In this work we explored a new track of applied mathematics for non-repayment risk modeling in microfinance. The results found are not exhaustive and the present model reflects the dynamics of the microcredit system and describes the interactions between the MFI and its borrowers and on the other hand with the borrowers themselves. The risk of non-repayment is deeply caused by asymmetric information leading to the failure of microcredit contracts, which is a real challenge for MFIs. Our model developed in this paper presents a decision support tool for MFIs, regardless of type of contacts, which allows it to well redirect their development strategy to minimize the risk of unpaid, in terms of,1) create new local agency or withdraw other evaluating over time the number of solvent and insolvent borrowers ; 2) target populations that are not yet integrated into the microfinance system and strengthen financial inclusion and 3) provide a mechanism for sustainable development to people who need to improve their living conditions.

To go further one can approach the analysis of the system to be able to understand and predict the behavior of the system and its performance as a function of the variation of the inputs of the model. Thus a control of the outlets that allows the MFI to mitigate risky portfolios and ensure its sustainability.

The Proposed model enhancement may result in still eligible results by MFIs based on new work in the theory of systems of nonlinear differential equations by study of the stability of the system and the associated bifurcation. Calibration of the model requires a punctual estimate or by confidence interval, model parameters by empirical study in areas where the MFI would like to settle or create local agencies with borrowers. Our contribution presents one of those works whose approach is new and which would be a possible trail to explore for the future.

## References

Aboulaich, R., Khamlichi, F., & Kaicer, M. (2013). Microcredit Scoring with Fuzzy Logic Model. *International Journal Statistics & Economics*, (1),109-117.

Armendariz, D., & Morduch, J. (2010). *The economics of micorfinance.* Cambridge: MIT press.

Besley, T., & Coate, S. (1995). Group lending, repayment incentives and social collateral. *Group lending, Journal of Development Economics*, (46),1-18.

Bhole, B., & Ogden, S. (2010). Group lending and individual lending with strategic default. *Journal of Development Economics*, (91), 348–363.

Bodie, Z., Merton, R. C., & Samuelson, W. F. (1992). Labor supply flexibility and portfolio choice in a life cycle model. *Journal of Economic Dynamics and Control, 16*(3-4), pp. 427-49.

Bourjade, S., & Schindele, I. (2012). Group lending with endogenous group size. *Economics Letters*, (117), 556-560.

Chłoń-Domińczak, A., Franco, D., & Palmer, E. (2012). The First Wave of NDC Reforms: The Experiences of Italy, Latvia, Poland,and Sweden. In R.

Holzmann, E. Palmer, & D. Robalino, *Nonfinancial Defined Contribution Pension Schemes in a Changing Pension World : Volume 1. Progress, Lessons, and Implementation* (pp. 31-84). Washington, DC: World Bank.

European Comission. (2015). *The 2015 Ageing Report. Economic and budgetary projections for the 28 EU Member States (2013-2060).* Brussels: European Commission. Directorate-General for Economic and Financial Affairs.

Galouchko, K., & Kucukreisoglu, L. (2016, 10 18). *bloomberg.com*. Retrieved 03 25, 2017, from https://www.bloomberg.com/news/articles/2016-10-18/poland-seeks-to-lock-in-low-yields-with-first-30-year-eurobond

Ghatak, M. (1999). Group lending, local information and peer selection. *Journal of Development Economics*, (60), 27–50.

Hulme, D., & Mosley, P. (1996). Financial Sustainability, Targeting the Poorest, and Income. *Focus,*, (5).

Kaddar, A., Sahbani, S., & Allaoui, h. T. (2018). Fluctuations in a delayed pegi model with economic characteristics of population. *Journal of nonlinear systems and applications*, 52–56.

Kaicer. (2014). *Decision making tools in microcredit.* Rabat: Mohammadia Engineering School.

Kaicer, M. (2014). *Decision making tools in microcredit.* Rabat: Mohammadia Engineering School.

Kaicer, M., & Aboulaich, R. (2014). Econometrics analysis of the failure in group lending. *International of innovation and applied studies,*, (5), 106-114.

Kaicer, M., & Aboulaich, R. (2014). Econometrics analysis of the failure in group lending. *Interna-tional of innovation and applied studies*, (5), 106-114.

Kermack, W. O., & McKendrick, A. G. (1927.). A contribution to the mathematical theory of epidemics. *Pro Royal Soc*, (115), 700-721.

Leiber, T. (1997). On the impact deterministic chaos on modern science and phylosophy of science: Implication for the phylosophy of technology. *Proceeding of a meeting,* (pp. N2, 23-50). Karlsruhe: The international academy of the phylosophy of science.

Maskin, E. (2011). Nash equilibrium and mechanism design. *Games and Economic Behavior,*, (71), 9–11.

Milgram, B. L. (2001). Operationalizing Microfinance: Women and Craftwork in Ifugao, Upland Philippines. *Human Organization*, (60)3, 212-224.

Norell, D. (2001). How To Reduce Arrears In Microfinance Institutions. *Journal of Microfinance / ESR Review*, (3), 1-16.

Park, A., & Ren, C. (January 2001). Microfinance with Chinese Characteristics. *World Development*, (29),1.

Perron, J.-L. (2017). *Barometer of Microfinance.* Paris: Yunus Center.

Perron, J.-L. (2017). *Micorfinance Barometer: 8th Edition.* Paris: Yunus CEnter.

Roy, A. D. (1952). Safety First and the Holding of Assets. *Econometrica, 20*(3), pp. 431-49.

Tassel, E. V. (1999). Group lending under asymmetric information. *Journal of Development Economics*, 60(1), 3-25.

Tedeschi, G. (2006). An dynamic incentives make microfinance more flexible? *Journal of development economics*, 80(1), 84-105.

Treynor, J., & Mazuy, K. (1966). Can Mutual Funds Outguess the Market? *Harvard Business Review, 44*, pp. 131-136.

# Appendix

| TABLE 1. STATISTICS OF MICROFINANCE ENVIRONMENT FROM THE WORLD BANK 2016 | | | | | |
|---|---|---|---|---|---|
| Areas | Number of borrowers (Million) | Annual growth | Share of rural borrowers | Portfolio (Milliard USD) | Number of MFIs |
| MENA | 2,4 | 3,2% | 9% | 1,4 | 30 |
| Eastern Europe and Central Asia | 3,1 | -11,1% | 67,7% | 9,3 | 151 |
| LATIN AMERICA AND THE CARIBBEAN | 23,2 | 8,1% | 39,8% | 42,5 | 355 |
| AFRICA | 7,2 | -0,6% | 70,8% | 8,7 | 211 |
| SOUTH ASIA | 78,3 | 23,5% | 68,8% | 23,5 | 222 |
| East Asia and Pacific | 17,8 | 9,2% | 43,2% | 16,5 | 148 |

SOURCE: DATA FROM STATISTICAL DATA BANK, THE WB, 2019
NOTE: SEE BAROMETRE OF MICORFINANCE (Perron, Micorfinance Barometer: 8th Edition, 2017)

FIGURE 1. INFORMATION FLOW IN MICROFINANCE ENVIRONMENT AS AREA, PORTFOLIO.

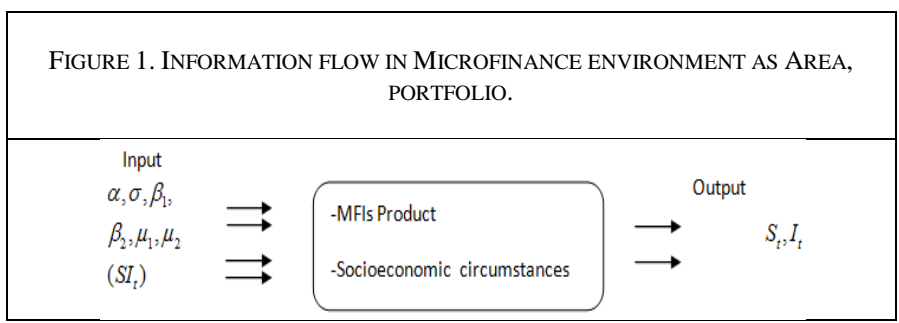

Input: $\alpha, \sigma, \beta_1, \beta_2, \mu_1, \mu_2$ $(SI_t)$ → -MFIs Product / -Socioeconomic circumstances → Output: $S_t, I_t$

Figure 2. CONCORDANCE BETWEEN THE TWO NUMBERS OF INSOLVENT AND INSOLVENT BORROWERS

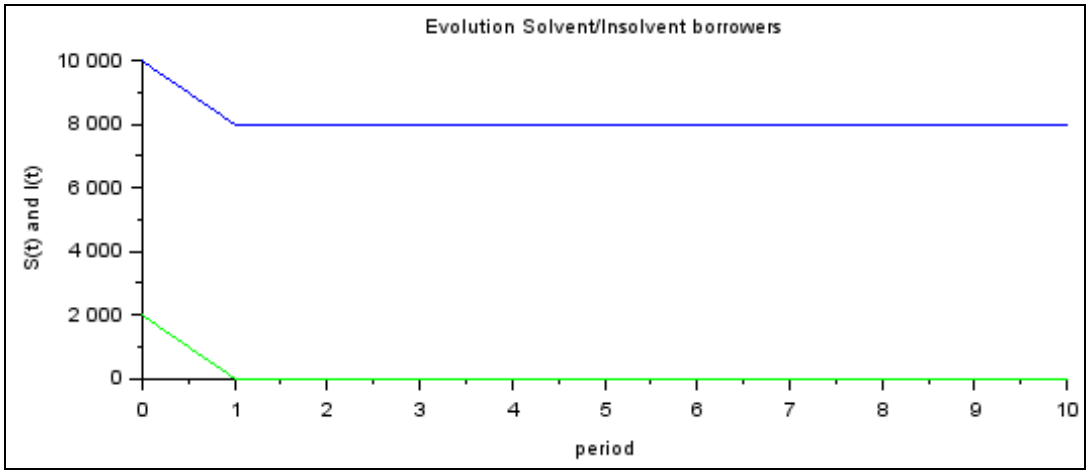

Figure 3. SITUATION EXPECTED BY AN MFI

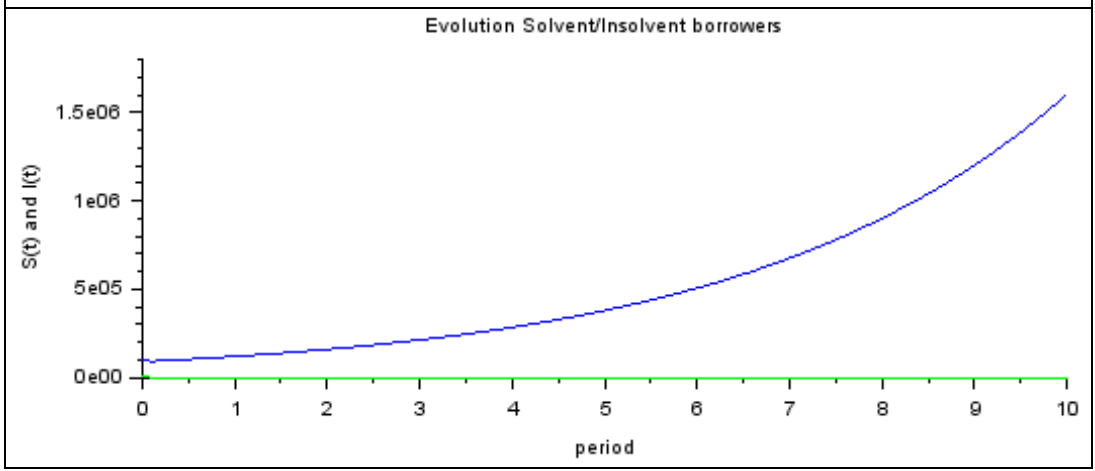

Figure 4. UNDESIRABLE SITUATION FOR MFI

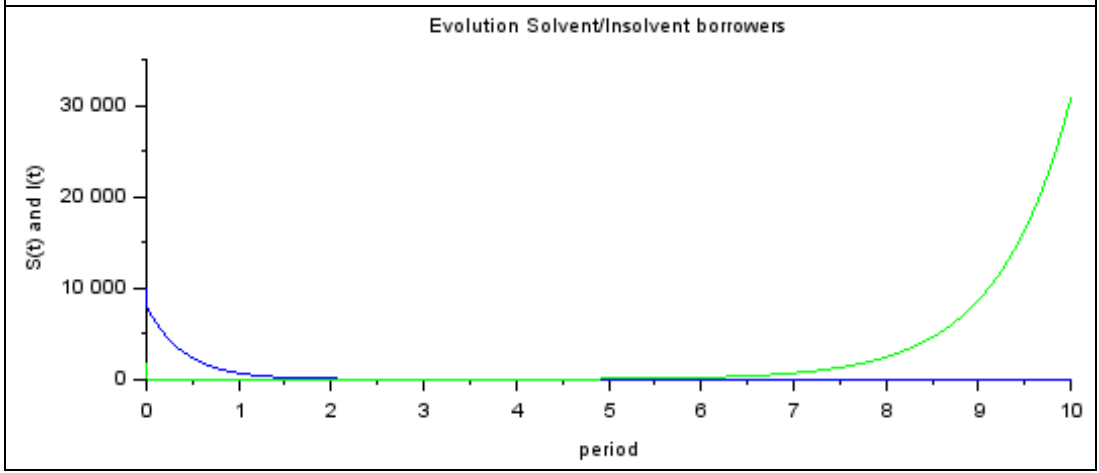

Figure 5. SITUATION WHERE THE MFI MUST CHANGE MANAGEMNT STRATEGY AS SOON AS POSSIBLE

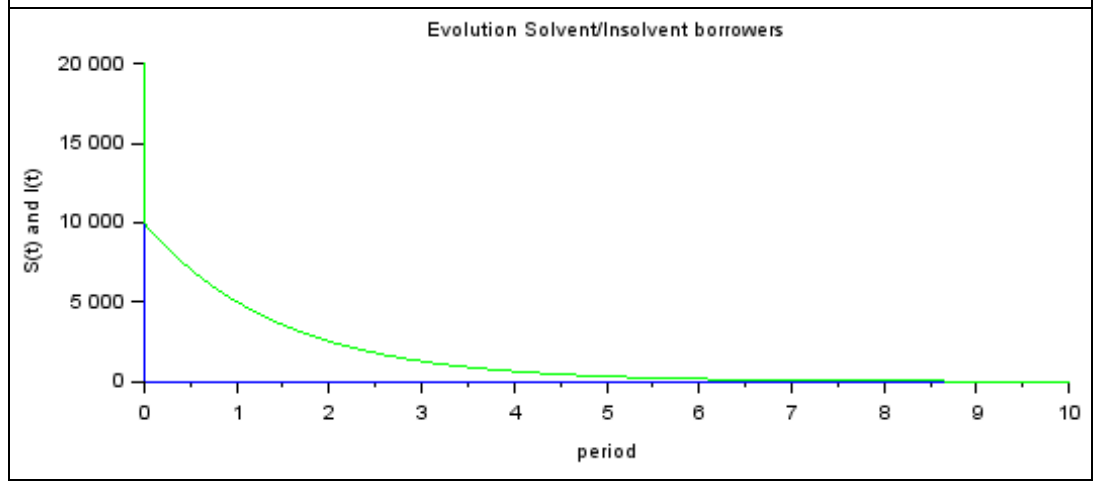